\documentclass[aps,twocolumn,pra,superscriptaddress,amsmath,showpacs,tightenlines]{revtex4}
\usepackage{epsfig,graphicx,times}
\usepackage{amstext}
\usepackage{amsmath}            
\usepackage{amssymb}            
\usepackage{graphicx}           
\usepackage{latexsym}
\usepackage{color}
\usepackage{bm}

\def \etal{\textit{et al.}}
\def\Title#1{\emph{#1}}

\begin{document}

\title{Tunable photon blockade in a hybrid system consisting of an optomechanical device coupled to a two-level system}

\author{Hui Wang}
\affiliation{Institute of Microelectronics, Tsinghua University,
Beijing 100084, China}

\author{Xiu Gu}
\affiliation{Institute of Microelectronics, Tsinghua University,
Beijing 100084, China}

\author{Yu-xi Liu}
\email{yuxiliu@mail.tsinghua.edu.cn} \affiliation{Institute of
Microelectronics, Tsinghua University, Beijing 100084, China}
\affiliation{Tsinghua National Laboratory for Information Science
and Technology (TNList), Beijing 100084, China} \affiliation{CEMS,
RIKEN, Saitama 351-0198, Japan}

\author{Adam Miranowicz}
\affiliation{Faculty of Physics, Adam Mickiewicz University,
61-614 Pozna\'n, Poland} \affiliation{CEMS, RIKEN, Saitama
351-0198, Japan}

\author{Franco Nori}
\affiliation{CEMS, RIKEN, Saitama 351-0198, Japan}
\affiliation{Physics Department, The University of Michigan, Ann
Arbor, Michigan 48109-1040, USA}

\date{\today}


\begin{abstract}
We study photon blockade and anti-bunching in the cavity of an
optomechanical system in which the mechanical resonator is coupled
to a two-level system (TLS).  In particular, we analyze the
effects of the coupling strength (to the mechanical mode),
transition frequency, and decay rate of TLS on the photon
blockade. The statistical properties of the cavity field are
affected by the TLS, because the TLS changes the energy-level
structure of the optomechanical system via dressed states formed
by the TLS and the mechanical resonator. We find that the photon
blockade and tunneling can be significantly changed by the
transition frequency of the TLS and the coupling strength between
the TLS and the mechanical resonator. Therefore, our study
provides a method to tune the photon blockade and tunneling using
a controllable TLS.
\end{abstract}

\maketitle

\section{Introduction}

Cavity optomechanics has attracted extensive theoretical and
experimental research activity in the last
decade~\cite{Marquardt,Aspelmeyer,PR1,PR2,review1,Schwab1,Schwab2,
Painter,Meystre,Tang1,Tang2,NatureMilestones,Xin}. It ranges from
testing fundamental aspects of quantum physics and gravity to
applications in quantum engineering, quantum
measurements~\cite{Braginsky92} and weak-force
detection~\cite{Caves,Bocko96,Buks06}. For example,
experiments~\cite{OConnell10} have demonstrated the quantum ground
state and single-phonon control in a mechanical resonator, which
is coupled to a superconducting TLS. It has also been shown that
the mechanical resonator can be used for frequency
conversion~\cite{Lin11, Hill, Bochmann, Lin12, Lin13}. By
controlling the frequency and the time intervals of a pumping
field, nonclassical states of the mechanical motion can be
prepared by carrier and sideband transition
processes~\cite{xu-phonon1,xu-phonon2,Long}.

\begin{figure}
\includegraphics[bb=10 360 360 770, width=6.3 cm, clip]{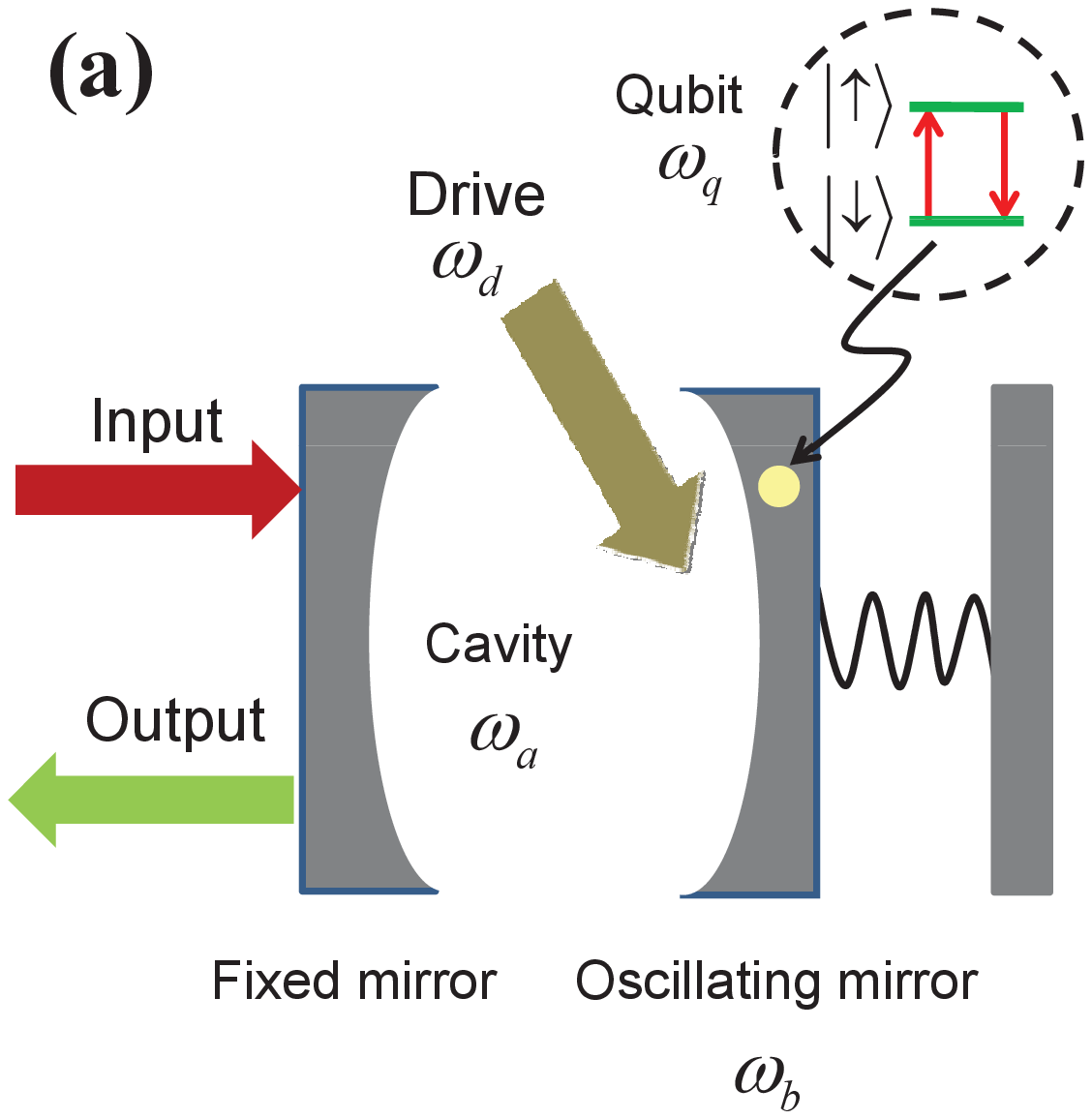}\\
\includegraphics[bb=-10 305 370 770, width=6.2 cm, clip]{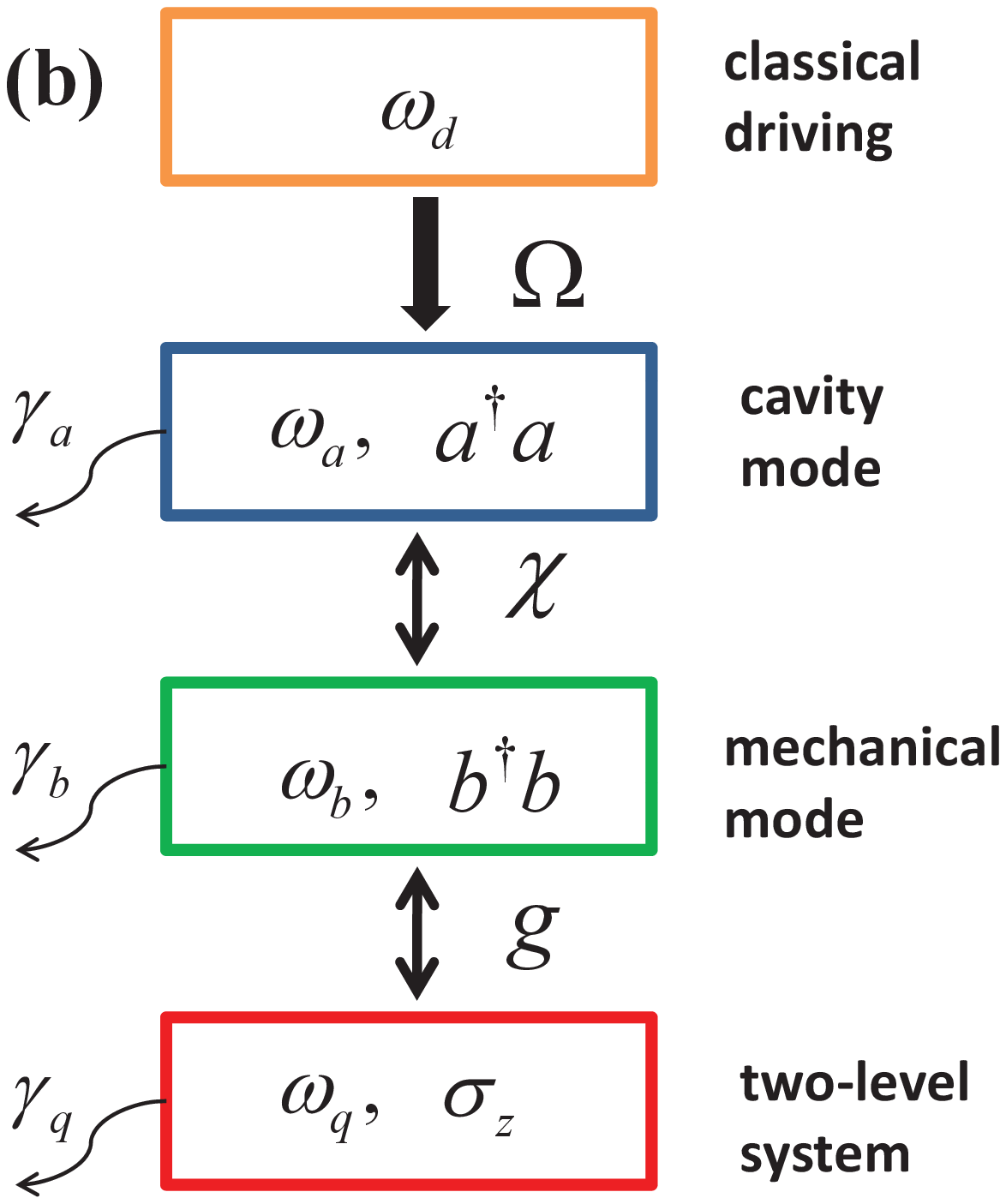}
\caption{(Color online) (a) Schematic diagram of a hybrid
structure consisting of a TLS coupled to the mechanical resonator
of an optomechanical system. The TLS (within the black dashed
circle) is denoted by a yellow dot inside the oscillating mirror
represented by a black spring. Here, $|\downarrow\rangle$ and
$|\uparrow\rangle $ denote the ground state and excited state,
respectively, of the TLS. The parameters $\omega_{q}$,
$\omega_{a}$, $\omega_{b}$, and $\omega_{d}$ denote the
frequencies of the TLS, cavity field, oscillating mirror, and
driving field, respectively. (b) A schematic diagram for the
couplings in the hybrid system with dissipation. The TLS is
coupled to the mechanical resonator by the Rabi-type with the
coupling strength $g$, the mechanical resonator is coupled to the
cavity field with coupling strength $\chi$, and the cavity field
is driven by an external field with amplitude $\Omega$. Here, $a$
($a^{\dagger}$) and $b$ ($b^{\dagger}$) are the annihilation
(creation) operators of the cavity mode and mechanical resonator,
respectively, and $\sigma_{x}=\sigma_{+}+\sigma_{-}$.
$\gamma_{a}$, $\gamma_{b}$, and $\gamma_{q}$ denote the decay
rates of the cavity field, the mechanical resonator and the TLS,
respectively.}\label{fig1}
\end{figure}

It is known that photon control can be realized in optomechanical
systems via an analogue of electromagnetically induced
transparency (EIT), well-known in quantum optics. For instance, it
has been found that EIT and photon scattering can be used to tune
photon transmission in optomechanical
systems~\cite{Agarwal,Weis,Liao1,Ma,Qi,Akram3}. A TLS coupled to
the cavity field of an optomechanical system can affect the photon
transmission and lead to nonclassical effects for the cavity
field~\cite{Pirkkalainen, Ciuti,Jia,Akram1,Akram2}. When the TLS
is a controllable superconducting qubit, we find that the EIT
window of the optomechanical system can be changed by the
superconducting qubit, or, in other words, that the mechanical
resonator can affect the absorption and dispersion of the circuit
QED system~\cite{hui}. Moreover, the mechanical resonator of an
optomechanical system can also interact with a TLS~\cite{ZLXiang},
which can affect the ground state cooling of the mechanical
resonator~\cite{Tdefects}, the nonlinearity of the cavity
field~\cite{hybrid2}, and so on. When the cavity field  in such a
hybrid system is driven by a strong classical field and a weak
probe field, the splitting of the phonon energy levels leads to
two-color EIT windows~\cite{Wang}, which can be switched to a
single one by adjusting the transition frequency of the TLS.

Photon control in an optomechanical system can also be realized
via photon blockade and tunneling, which result from the
nonlinearity of the cavity field. Photon blockade prevents
subsequent photons from resonantly entering the cavity, while the
photon-induced tunneling increases the probability of subsequent
photons entering the cavity. Thus, photon blockade corresponds to
a single-photon transition process, while photon tunneling
corresponds to two-photon or multi-photon transition processes. If
an optomechanical system coupled to a TLS via a mechanical
resonator, both the mechanical resonator and the TLS can induce a
nonlinearity in the cavity field, and thus they can be used to
realize photon blockade and tunneling. Photon blockade has been
studied in various systems, e.g., cavity
QED~\cite{Tian92,Leonski94, Miran96, Imamoglu97, Rebic99, Kim99,
Rebic02, Smolyaninov02,Birnbaum05}, circuit
QED~\cite{Hoffman11,Lang11,Didier11,Liu14}, and optomechanical
devices~\cite{Rabl,Nunnenkamp11,Liao13}. In addition to the
single-photon blockade, multiphoton blockade was also studied
theoretically (see,
e.g.,~\cite{Adam13,Adam14a,Adam14b,Hovsepyan14} and references
therein) and even observed experimentally~\cite{Smolyaninov02,
Faraon08,Majumdar12,Schuster08,Kubanek08}. However, to our
knowledge, there is no study on how to control photon blockade and
tunneling.

In this paper, we study a method to tune photon blockade and
anti-bunching in an optomechanical system  via a TLS which is
coupled to the mechanical resonator of the optomechanical device.
In such a hybrid system~\cite{hybrid2},  the dressed states formed
by the mechanical resonator and the TLS affect the photon and
phonon blockade of the optomechanical system. It is known that the
eigenstates of phonons in an optomechanical devices are described
by displaced Fock states~\cite{Rabl} due to the phonon-photon
coupling via the radiation pressure. Therefore,  the dressed
states in the hybrid system should be more complicated, because
they formed by the displaced phonon states of the mechanical
resonator and the TLS~\cite{alsing1,Kim}. If the mechanical mode
and the TLS are in the ultrastrong coupling regime, the rotating
wave approximation (RWA) doesn't work, the  Rabi type interaction
should be considered. In particular, the effect of strong and
ultrastrong coupling on the photon blockade is analyzed in many
systems~\cite{Cao,Zueco,Niemczyk,Crisp}.

The model studied here is a combination of the usual prototype
optomechanical models. Hybrid systems composed of a TLS coupled to
the cavity field of an optomechanical system have been studied
widely (see, e.g., the recent Refs.~\cite{Pirkkalainen,
Ciuti,Jia,Akram1,Akram2,Wang} and references therein).
Specifically, we consider a standard Hamiltonian for two
interacting oscillators (i.e., optical and mechanical resonators)
in which the mechanical oscillator interacts also with a two-level
system (TLS). It is worth noting that the model studied here is
nontrivial because the couplings between its constituent
subsystems are nonlinear. For example, the interaction between the
two oscillators is proportional to the photon number and the
position of a mechanical resonator. This nonlinear interactions
can induce nonlinearity of the oscillators. For example, as will
be shown below, the optical oscillator, due to its interaction
with the mechanical oscillator, can be effectively described by a
Kerr-type nonlinearity. It is known that the standard Kerr
nonlinearity can induce various nonclassical
effects~\cite{Tanas03,HarocheBook}. These include
self-squeezing~\cite{Tanas83,Milburn86,Yamamoto86,Tanas91},
generation of two-component~\cite{Yurke86,Milburn86,Tombesi87} and
multi-component~\cite{Miran90,Kirchmair13} Schr\"odinger cat
states, and photon antibunching (if the nonlinearity is driven).
The latter is a signature of photon blockade (also referred to as
optical state truncation)~\cite{Tian92,Leonski94,Imamoglu97}, as
also studied here.

The creation of photons due to the mechanical resonator (i.e.,
oscillating mirror, which causes time-dependent variations of the
geometry of our mesoscopic optomechanical system) can be
interpreted as a result of the dynamical Casimir effect (DCE),
which is also known as non-stationary Casimir effect or
motion-induced radiation (from a dynamically deforming mirror). As
explained in Ref.~\cite{Dodonov11}: ``The term `dynamical Casimir
effect' is used nowadays for a rather wide group of phenomena
whose common feature is the creation of quanta (photons) from the
initial vacuum (or some other) state of some field
(electromagnetic field in the majority of cases) due to
time-dependent variations of the geometry (dimensions) or material
properties (e.g., the dielectric constant or conductivity) of some
macroscopic system.'' Specifically, we can interpret the
occurrence of photon blockade in the studied system as follows: As
mentioned above, the nonlinear interaction between the mechanical
and optical resonators of our system can induce an effective
Kerr-type nonlinearity of the optical resonator. This driven Kerr
nonlinearity can result in photon blockade. Note that this driving
is applied directly via the coupling of the mechanical and optical
resonators (being related to the DCE) and indirectly via the
coupling of the mechanical resonator with the TLS.

The DCE was studied in analogous systems in a number of recent
works (see, e.g.,
Refs.~\cite{Dodonov10,J1,J2,Wilson11,Dodonov11,Dodonov12a,Dodonov12b}).
In particular, Ref.~\cite{Dodonov11} analyzed strong modifications
of the cavity field statistics in the DCE due to the interaction
with TLSs. Here, we study the effect of a single TLS on photon
blockade. The light generated via the DCE can exhibit various
nonclassical properties~\cite{Dodonov10} including
squeezing~\cite{J1,J2,Wilson11,Dodonov97,Dodonov00}.

The paper is organized as follows: In Sec.~II, we describe the
theoretical model. In Sec.~III, we write down the master equation
and derive the analytical solution  in the weak-pumping limit. The
photon blockade is analyzed via the second-order degree of
coherence in Sec.~IV. We finally summarize our results in Sec.~V.

\section{Energy level structure of the hybrid system}\label{model}

\subsection{Theoretical model}

As schematically shown in Fig.~\ref{fig1}, we study a hybrid
system which consists of an optomechanical cavity coupled to  a
TLS with its mechanical mode.  We assume that there is no direct
coupling between the TLS and the cavity field of the
optomechanical part. In this case, the Hamiltonian of the hybrid
system can be written as
\begin{eqnarray}\label{eq:1}
H_{0}&=&\hbar\omega_{a}a^{\dag}a+\hbar\omega_{b}b^{\dagger}b
+\frac{\hbar }{2}\omega_{q}\sigma_{z}-\hbar\chi a^{\dag}a \left(b^{\dagger}+b\right)\nonumber\\
& &+\hbar g\left(b^{\dagger} +b\right)\sigma_{x}.
\end{eqnarray}
Here, $a$ ($a^{\dagger}$) and $b$ ($b^{\dagger}$) are the
annihilation (creation) operators of the cavity field and the
mechanical resonator, respectively. The frequencies of the cavity
field and the mechanical resonator are denoted by $\omega_{a}$ and
$\omega_{b}$, respectively. The transition frequency of the TLS is
$\omega_{q}$.  The Pauli operators $\sigma_{z}$ and $\sigma_{x}$
are used to describe the TLS with the ladder operators $\sigma_{\pm}$ defined by
$\sigma_{x}=\sigma_{+}+\sigma_{-}$. The coupling strength between
the cavity field and the mechanical resonator is $\chi$, and the
parameter $g$ describes the coupling strength between the
mechanical resonator and the TLS.

\begin{figure}
\includegraphics[bb=-10 290 600 820, width=8.4 cm, clip]{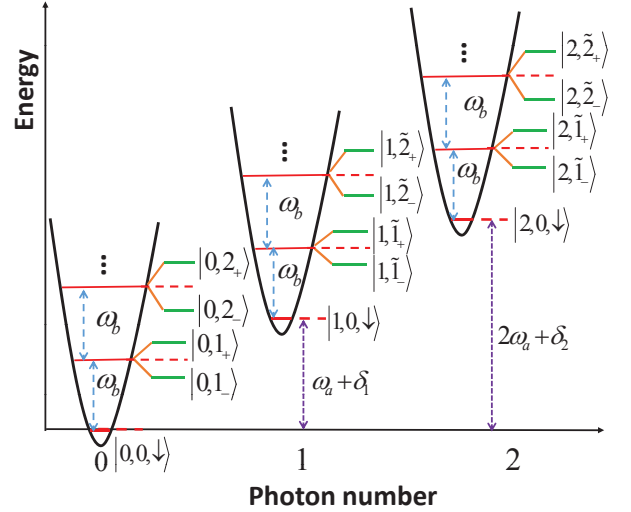}
\caption{(Color online) Schematic diagram of the energy levels of
the hybrid system when the TLS resonantly interacts with the
mechanical resonator  (the ground state energy of the TLS is
assumed to be $0$). Here $|n\rangle$ ($n=0,1,2,\cdot\cdot\cdot$)
represent the Fock states of the photons. If the photon number is
zero, the dressed states of the phonon and the TLS are:
$|1_{\pm}\rangle=|1,\downarrow\rangle \pm |0,\uparrow\rangle$ and
$|2_{\pm}\rangle=|2,\downarrow\rangle \pm |1,\uparrow\rangle$.
Here $|\uparrow\rangle$ and $|\downarrow\rangle$ denote the
eigenstates of the TLS which is not coupled to mechanical
resonator. When the photon number is nonzero,  the expressions of
the wave functions for the dressed states are given by
$|\tilde{m}_{\pm}(n)\rangle$ in Eq.~(\ref{eq:14})
($n,m=0,1,2,\cdot\cdot\cdot$), and the corresponding eigenvalues
are given in Eq.~(\ref{eq:13}) (in the stable regime). Here
$\delta_{1}=\Delta^{\prime}(1)-\Delta_{0}$ and
$\delta_{2}=\Delta^{\prime}(2)-4\Delta_{0}$, where
$\Delta^{\prime}(n)=2g\chi\langle \sigma_{x}\rangle/\omega_{b}$
and $\Delta_{0}=\chi^{2}/\omega_{b}$. }\label{fig2}
\end{figure}

It has been shown (e.g., in Ref.~\cite{xu-phonon1}) that the
mechanical resonator can mediate a Kerr nonlinear interaction
between photons of the cavity field in an optomechanical system.
Thus, when  the mechanical resonator is coupled to the TLS,  the
mechanical resonator  will induce the effect of the TLS on the
nonlinearity of the cavity field.  To see this clearly, we apply a
unitary transformation, $U=\exp{[-\chi a^{\dagger}a
(b^{\dagger}-b)/\omega_{b}]}$, to Eq.~(\ref{eq:1}).  Then the
total Hamiltonian in Eq.~(\ref{eq:1}) becomes
\begin{eqnarray}\label{eq:2}
H^{\prime}_{0}&=&\hbar\omega_{a}a^{\dag}a+\hbar\frac{2
g\chi}{\omega_{b}}\sigma_{x}a^{\dag}a+\frac{\hbar
}{2}\omega_{q}\sigma_{z}+\hbar\omega_{b}b^{\dagger}b
\nonumber\\
& &-\hbar\frac{\chi^{2}}{\omega_{b}} a^{\dagger}a
a^{\dagger}a+\hbar g\left(b^{\dagger} +b\right)\sigma_{x}.
\end{eqnarray}
Besides the energy level shift $-n^{2}\chi^2/\omega_{b}$
($n=0,1,2,\cdot\cdot\cdot$) induced by the mechanical resonator
with the photon number $n$, the interaction between the TLS and
the cavity field through $\hbar 2
g\chi\sigma_{x}a^{\dag}a/\omega_{b}$ also leads to a photon energy
level shift $2 g\chi \langle\sigma_{x}\rangle/\omega_{b}$, which
is twice the result found with the RWA~\cite{Wang}. This
interaction induces a new nonlinearity of the cavity field. For
example, in the case of large detuning between the cavity field
and the TLS, the TLS can induce another photon-photon Kerr
interaction term~\cite{liuphonon}. If we define
$\Delta_{0}=\chi^{2}/\omega_{b}$ as the photon nonlinearity
induced by the mechanical resonator in the optomechanical system,
then the total photon energy levels shifts for the one-photon and
two-photon states of the hybrid system are
$\delta_{1}=\Delta^{\prime}(1)-\Delta_{0}$, and
$\delta_{2}=\Delta^{\prime}(2)-4\Delta_{0}$, respectively, as
schematically shown in Fig.~\ref{fig2}. Here, both
$\Delta^{\prime}(n)=2g\chi\langle \sigma_{x}\rangle/\omega_{b}$
and $\langle \sigma_{x}\rangle$ depend on the photon number
$n$~\cite{Wang}.

\subsection{Eigenvalues and eigenstates}
We now analyze the eigenvalues and eigenstates of the system when
the cavity field of the optomechanical system is in a Fock state
$|n\rangle$ with the photon number $n$. In this case, the quantity
$\chi n$ can be considered as an effective driving field for the
coupled system of the mechanical resonator and the TLS, and the
effective Hamiltonian of the mechanical resonator and the TLS for
the photon Fock state $|n\rangle$ can be given, from Eq.~(\ref{eq:1}), as
\begin{eqnarray}\label{eq:12}
H_{b}=\hbar\omega_{b}b^{\dagger}b+\frac{\hbar
}{2}\omega_{q}\sigma_{z}+\hbar g\left(b^{\dagger}
+b\right)\sigma_{x}-\hbar\chi n\left(b^{\dagger}+b\right),
\end{eqnarray}
where the constant term $\hbar\omega_{a}n$ has been neglected.

Let us first study the eigenvalues and eigenstates when the
mechanical resonator and the TLS satisfy the resonant interaction
condition in Eq.~(\ref{eq:13}), i.e.
$\Delta_{d}=\omega_{b}-\omega_{q}=0$.  Under the RWA, the dressed
state energy levels in the interaction picture can be given
as~\cite{alsing1},
\begin{eqnarray}
E_{n,m,\pm}&=&\pm\hbar\sqrt{m}g\left[1-\left(\frac{2\chi
n}{g}\right)^{2}\right]^{3/4}.\label{eq:13}
\end{eqnarray}
Here, $m$ ($m=1,2,\ldots$) denotes the phonon number. If the
effective driving field is not very strong, that is, $2\chi n<g$,
the dressed states will be stable. Otherwise the phonons have
large chances to transit to high energy levels and the dressed
states will be unstable~\cite{alsing1,Kim}. The energy levels of
the dressed states are also functions of the photon number. The
dressed states are always stable for the $0$ photon state,
however, the dressed states become unstable when $n> g/(2\chi)$.
We can see from Eq.~(\ref{eq:13}) that each energy level has an
extra term compared with $\pm\hbar\sqrt{m}g$ of the common dressed
states in the resonant interaction between the TLS and the mechanical resonator.
 Here, the splitting width of the dressed states is
affected by the quantum states of both photons and phonons.

If the photon number is zero, the eigenvalues corresponding to the
dressed states in  Eq.~(\ref{eq:13}) become $\pm \hbar \sqrt{m}g$,
and the  corresponding dressed states can be written as
$|m_{\pm}\rangle=\left[|m,\downarrow\rangle\pm
|m-1,\uparrow\rangle\right]/\sqrt{2}$, which are the common
dressed states of the resonant interaction between the TLS and the mechanical resonator,
as schematically shown in Fig.~\ref{fig2}. The
dressed states, corresponding to the eigenvalues in
Eq.~(\ref{eq:13}), can be written as~\cite{alsing1},
\begin{eqnarray}
|\tilde{m}_{\pm}(n)\rangle &=&\frac{1}{\sqrt{2}}\left[|\eta,\beta(E_{n,m,\pm});m-1\rangle|P\rangle\right.\nonumber\\
& &\left.\pm
i|\eta,\beta(E_{n,m,\pm});m\rangle|M\rangle\right].\label{eq:14}
\end{eqnarray}
Here $|P\rangle$ and $|M\rangle$ correspond to quantum states of
the TLS, and the expression of $|P\rangle$ is given by
\begin{eqnarray}
|P\rangle&=&\frac{1}{\sqrt{2}}\left[\left(1+\sqrt{\varepsilon}\right)^{1/2}|\uparrow\rangle
-\left(1-\sqrt{\varepsilon}\right)^{1/2}|\downarrow\rangle\right],
\label{eq:15}
\end{eqnarray}
with $\varepsilon=1-\left(2\chi n/g\right)^{2}$. The expression of
$|M\rangle$ can be obtained by replacing $|\uparrow\rangle$ and
$|\downarrow\rangle$ with $|\downarrow\rangle$ and
$|\uparrow\rangle $ in Eq.~(\ref{eq:15}). The state
$|\eta;\beta;m\rangle$ in Eq.~(\ref{eq:14}) is given as
\begin{eqnarray}
|\eta;\beta;m\rangle=D(\beta)S(\eta)|m\rangle.\label{eq:16}
\end{eqnarray}
The squeezing operator in Eq.~(\ref{eq:16}) is defined as
$S(\eta)=\exp{[\frac{1}{2}(\eta b^{\dagger 2}-\eta^{\ast}
b^{2})]}$, while the expression of the displacement operator is
$D(\beta)=\exp{(\beta b^{\dagger}-\beta^{\ast} b)}$. The
parameters $\beta$ and $\eta$ are defined as $\beta(E)=2i\chi n
E/\left(\hbar g^{2}\varepsilon\right)$, $\eta=r$, and
$\exp{(2r)}=\sqrt{\varepsilon}$. We can find that the dressed
states in the hybrid system are formed by the superposition states
$|P\rangle$ and $|M\rangle$ of the TLS, not the eigenstates
$|\uparrow\rangle$ or $|\downarrow\rangle$ as in common dressed
states. This leads to more complicated phenomena when a probe
field passes through such a system. If the mechanical resonator
and the TLS are in the large detuning regime, the energy level
spacing between two dressed states becomes larger~\cite{alsing1}.

In circuit QED, the standard photon blockade is significantly
changed by the ultrastrong coupling between the cavity field and
the TLS~\cite{Hartmann}. Since the sideband-transition processes
in optomechanics usually accompany the absorption or emission of
phonons, the variation of phonon energy levels in a hybrid system
can also affect transitions of photons. So we will also study the
effect of the ultrastrong coupling between the mechanical
resonator and the TLS on the photon blockade in the hybrid
devices.

\section{Master Equation and weak pumping limit}\label{Master}

\subsection{Master equation}

To study photon blockade, we assume that the cavity field of the
hybrid system is driven by a classical field with frequency
$\omega_{d}$, the coupling strength between the driving field and
the cavity field is $|\Omega|$. In the rotating reference frame at
the frequency $\omega_{d}$, the Hamiltonian of the driven hybrid
system becomes
\begin{eqnarray}\label{eq:3}
H_{r}&=&\hbar\Delta_{a} a^{\dag}a+\hbar\omega_{b}b^{\dagger}b+\frac{\hbar }{2}\omega_{q}\sigma_{z}-\hbar\chi a^{\dag}a \left(b^{\dagger}+b\right)\nonumber\\
& &+\hbar g\left(b^{\dagger}
+b\right)\sigma_{x}+i\hbar\left(\Omega
a^{\dag}-\Omega^{\ast}a\right),
\end{eqnarray}
where $\Delta_{a}=\omega_{a}-\omega_{d}$ describes the detuning
between the cavity field and the driving filed.

After introducing the environmental noise, the master equation of
the density operator $\rho$ for the driven hybrid system can be
given as
\begin{eqnarray}\label{eq:4}
\dot{\rho}=\frac{i}{\hbar}[\rho,
H_{r}]+L_{a}(\rho)+L_{b}(\rho)+L_{\sigma}(\rho).
\end{eqnarray}
The Lindblad dissipators for the photons and phonons are given by
\begin{eqnarray}
L_{o}(\rho)&=&\gamma_{o}n_{o}\left(o\rho o^{\dagger}+o^{\dagger}\rho o-o^{\dagger}o \rho-\rho o^{\dagger}o\right)\nonumber\\
& &+\frac{\gamma_{o}}{2}\left(2o\rho o^{\dagger}-o^{\dagger}o
\rho-\rho o^{\dagger}o\right),
\end{eqnarray}
where $o=a$ or $b$ corresponds to the variables of the photon or
phonon, respectively. The Lindblad dissipator for the two-level
system is
\begin{eqnarray}
L_{\sigma}(\rho)&=&\gamma_{q}n_{q}\left(\sigma_{-}\rho\sigma_{+}+\sigma_{+}\rho \sigma_{-}-\sigma_{+}\sigma_{-} \rho-\rho \sigma_{+}\sigma_{-}\right)\nonumber\\
& &+\frac{\gamma_{q}}{2}\left(2\sigma_{-}\rho
\sigma_{+}-\sigma_{+}\sigma_{-} \rho-\rho
\sigma_{+}\sigma_{-}\right).
\end{eqnarray}
This type of master equations was also studied by Kossakowski
\emph{et al.}~\cite{Kossakowski72, Kossakowski76, Kossakowski78}.
Here $\gamma_{a}$, $\gamma_{b}$, and $\gamma_{q}$ are the decay
rates of the photon, the phonon, and the TLS, respectively, while
$n_{a}$, $n_{b}$, and $n_{q}$ correspond to thermal fluctuation
quantum numbers, with $n_{i}=1/[\exp(\hbar\omega_{i}/(k_{B}T))]$
($i=a, b, q$) where  $k_{B}$ is the Boltzmann constant and $T$ is
the temperature. Usually, the thermal photon number $n_{a}$ can be
neglected in the low-temperature limit  because of the high
frequency of the cavity field.

The master equation in Eq.~(\ref{eq:4}) can also be numerically
solved in the complete basis $|n, m,z\rangle$ (for
$n,m=0,1,2,\ldots$, and $z=\uparrow,\downarrow$) in the case of
weak driving field and low temperatures~\cite{Tan,J1,J2}. Because
higher excited states can be neglected in this case, the photon
and phonon numbers can be truncated to small values. By
numerically solving the master equation, we can obtain $\rho$
which in turn lets us calculate various physical properties of the
hybrid system.

\subsection{Analytical solutions in the weak-driving limit}

If the driving field coupling $\Omega$ is very weak in comparison
to the Kerr nonlinearity, and also the temperature is very low,
then, due to photon blockade, only lower energy levels of the
cavity field and mechanical resonator are occupied. If the photon
number $n$ and phonon number $m$ are truncated to $n=2$ and $m=1$,
respectively, then the quantum state of the hybrid system can be
written by~\cite{xunwei2,Bamba,Savona}
\begin{eqnarray}\label{eq:5}
|\psi\rangle&=&C_{00\downarrow}|0,0,\downarrow\rangle+C_{00\uparrow}|0,0,\uparrow\rangle+C_{10\downarrow}|1,0,\downarrow\rangle\nonumber\\
& &+C_{10\uparrow}|1,0,\uparrow\rangle+C_{01\downarrow}|0,1,\downarrow\rangle+C_{01\uparrow}|0,1,\uparrow\rangle\nonumber\\
& &+C_{20\downarrow}|2,0,\downarrow\rangle+C_{20\uparrow}|2,0,\uparrow\rangle+C_{11\downarrow}|1,1,\downarrow\rangle\nonumber\\
&
&+C_{11\uparrow}|1,1,\uparrow\rangle+C_{21\downarrow}|2,1,\downarrow\rangle.
\end{eqnarray}
The coefficients $C_{nmk}$ (with photon numbers $n=0,1,2$, phonon
numbers $m=0,1$, and the eigenvalues $k=\downarrow,\uparrow$ of
the dressed TLS states) describe the amplitudes of the
corresponding quantum states, and $p_{nmk}=|C_{nmk}|^2$ are the
corresponding occupation probabilities.

We use the second-order degree of coherence to describe the
statistical properties  of the cavity field. The equal-time
second-order degree of coherence is defined by
\begin{eqnarray}\label{eq:14-1}
g^{(2)}(0)=\frac{\langle
a^{\dagger}(t)a^{\dagger}(t)a(t)a(t)\rangle}{\langle
a^{\dagger}(t)a(t)\rangle^{2}}.
\end{eqnarray}
In the weak-driving limit, using Eq.~(\ref{eq:5}) and
Eq.~(\ref{eq:14-1}), the second-order degree of coherence can be
approximately given as
\begin{eqnarray}\label{eq:9}
g^{(2)}(0)\approx\frac{2\left(|C_{20\downarrow}|^{2}+|C_{20\uparrow}|^{2}+|C_{21\downarrow}|^{2}\right)}{(|C_{10\downarrow}|^{2}+|C_{10\uparrow}|^{2}+|C_{11\downarrow}|^{2}+|C_{11\uparrow}|^{2})^{2}}.
\end{eqnarray}
The result of Eq.~(\ref{eq:9}) can be used to approximately
describe the photon statistical properties in the limit of weak
driving  and low temperatures. This will be compared with
numerical results, calculated using the master equation, in the
following sections.

To obtain the coefficients  $C_{10\downarrow}$, $C_{10\uparrow}$,
$C_{11\downarrow}$,  $C_{11\uparrow}$, $C_{20\downarrow}$,
$C_{20\uparrow}$, and $C_{21\downarrow}$ in Eq.~(\ref{eq:5}), we
solve the Schr\"{o}dinger equation for the quantum state
$|\psi\rangle$ of the hybrid system
\begin{equation}\label{eq:6}
i\frac{d|\psi\rangle}{dt}=H^{\prime}_{r}|\psi\rangle.
\end{equation}
Here, the effective non-Hermitian Hamiltonian
\begin{eqnarray}\label{eq:7}
H^{\prime}_{r}&=&\hbar\Delta^{\prime}_{a} a^{\dag}a+\hbar\omega^{\prime}_{m}b^{\dagger}b+\frac{\hbar }{2}\omega^{\prime}_{q}\sigma_{z}-\hbar\chi a^{\dag}a \left(b^{\dagger}+b\right)\nonumber\\
& &+\hbar g\left(b^{\dagger}
+b\right)\sigma_{x}+i\hbar\left(\Omega
a^{\dag}-\Omega^{\ast}a\right),
\end{eqnarray}
includes dissipations with $\Delta^{\prime}_{a}=
\Delta_{a}-i\gamma_{a}/2$,
$\omega^{\prime}_{b}=\omega_{b}-i\gamma_{b}/2$, and
$\omega^{\prime}_{q}=\omega_{q}-i\gamma_{q}/2$. Here we assume
that the thermal fluctuation of the photons, phonons and the TLS
can be neglected in the extreme low-temperature limit.

Because we are interested in the statistical properties of the
cavity field in the steady state, thus we can set
$d|\psi\rangle/dt=0$. By substituting Eqs.~(\ref{eq:5})
and~(\ref{eq:7}) into Eq.~(\ref{eq:6}),  we can obtain linear
equations, as shown in Eqs.~(\ref{eq:A1})-(\ref{eq:A10}) of the
Appendix~\ref{App-A}. By solving these linear equations, we can
obtain the coefficients in Eq.~(\ref{eq:5}), that is,
\begin{eqnarray}\label{eq:8}
C_{11\downarrow}=i\Omega\eta_{1},~~~C_{11\uparrow}=i\Omega\eta_{2},\\
C_{10\downarrow}=i\Omega\eta_{3},~~~
C_{10\uparrow}=i\Omega\eta_{4},\\
C_{20\downarrow}=\Omega^{2}\eta_{5},~~~
C_{20\uparrow}=\Omega^{2}\eta_{6},\\
C_{21\downarrow}=\Omega^{2}\eta_{7}.
\end{eqnarray}
The expressions of $\eta_{i} (i=1,2,\cdot\cdot\cdot,7)$ can be
found in Eqs.~(\ref{eq:A12}). In the weak-driving and
low-temperature limit, we find that $C_{10\downarrow}$,
$C_{10\uparrow}$, $C_{11\downarrow}$, and $C_{11\uparrow}$ are
proportional to $\Omega$, while $C_{20\downarrow}$,
$C_{20\uparrow}$, and $C_{21\downarrow}$ are proportional to
$\Omega^{2}$. The value of $C_{00\downarrow}$ will be close to $1$
and the amplitudes of the excited state tend to $0$ if the value
of $\Omega\rightarrow 0$.

\section{Photon Blockade}\label{blockade}

\begin{figure}
\includegraphics[bb=0 190 570 650, width=8.65 cm, clip]{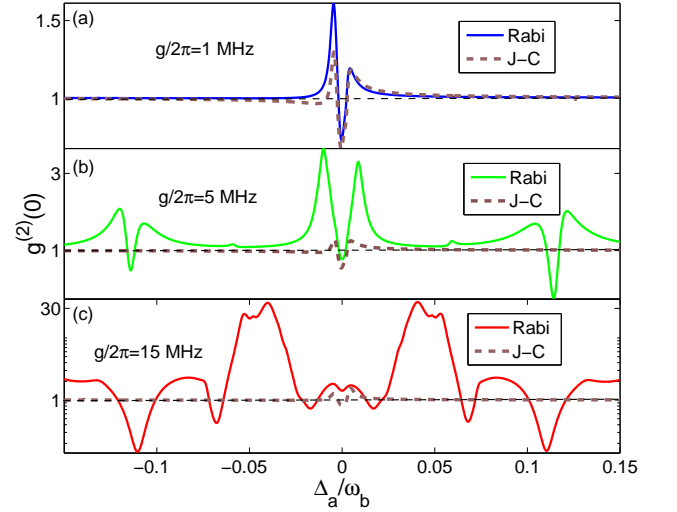}
\caption{(Color online) Equal-time second-order degree of
coherence $g^{(2)}(0)$ as a function of
$\Delta_{a}/\omega_{b}=(\omega_{a}-\omega_{d})/\omega_{b}$ in the cavity steady-state limit.
The solid curve in each panel is plotted with the Rabi model (without
RWA), while the dashed curve is plotted with the
Jaynes-Cummings(J-C) model (with RWA). The curves in the three
panels correspond to different coupling strengths between the
mechanical mode and the TLS: (a) $g/(2\pi)=1$ MHz; (b)
$g/(2\pi)=5$ MHz; and (c) $g/(2\pi)=15$ MHz. The other parameters
for the three solid curves are:
$\omega_{b}/(2\pi)=\omega_{q}/(2\pi)=10$ MHz,
$\gamma_{a}/(2\pi)=0.02$ MHz, $\gamma_{b}/(2\pi)=0.001$ MHz,
$\gamma_{q}/(2\pi)=0.002$ MHz, $\chi/(2\pi)=0.2$ MHz, $g/(2\pi)=4$
MHz, $T=1$ mK, and $|\Omega|/(2\pi)=0.02$ MHz. }\label{fig3}
\end{figure}

In an optomechanical system, the mechanical resonator leads to the
nonlinearity and energy levels shift of the cavity field. For the
single-photon state, such shift is
$\Delta_{0}=\chi^{2}/\omega_{b}$, while it is $4\Delta_{0}$ for
the two-photon state~\cite{Rabl}. Both the photon
blockade~\cite{Rabl} and tunneling~\cite{xunwei1,Roque} can occur
in the strong single-photon optomechanical coupling regime.
Besides mechanical mode, the TLS can also lead to the variation of
photon energy levels (see Eq.~(\ref{eq:2})). The energy level
structure of the hybrid system becomes very complex since
$\langle\sigma_{x}\rangle$ is a complicated function of
$\chi$~\cite{Wang}.  The dressed states formed by the TLS and the
mechanical mode lead to the splitting of phonon energy levels (in
standard optomechanical systems).

We now study how a TLS affects the photon blockade of a hybrid
system via the second-order degree of coherence
\begin{eqnarray}\label{eq:22-1}
g^{(2)}(0)=\frac{\text{Tr}(\rho a^{\dagger
2}a^{2})}{[\text{Tr}(\rho a^{\dagger }a)]^{2}},
\end{eqnarray}
which is calculated here using the master equation in
Eq.~(\ref{eq:4}) and will be compared to the result calculated
using Eq.~(\ref{eq:9}). The value of $g^{(2)}(0)<1 $
($g^{(2)}(0)>1$) corresponds to sub-Poisson (or super-Poisson)
statistics of the cavity field, which is a nonclassical
(classical) effect. This effect of the sub-Poisson photon statistics is often referred to as photon antibunching.
The dips ( resonant peaks) of $g^{(2)}(0)$ can
be used to characterize the photon blockade (tunneling)  processes.
The photon blockade describes the single-photon transition, while
the photon tunneling corresponds to a multi-photon resonant
transition.

We plot $g^{(2)}(0)$  as a function of $\Delta_{a}/\omega_{b}$ in
Fig.~\ref{fig3} by using the master equation in Eq.~(\ref{eq:4}),
the curves in different panels correspond to different values of
the coupling strength $g$ between the mechanical mode and the TLS.
To further study the effect of the counter-rotating term on the
photon blockade, using the master equation in Eq.~(\ref{eq:4}), we
compare the numerical results of $g^{(2)}(0)$ with (solid curves)
and without (dashed curves) the RWA. The minimum value of
$g^{(2)}(0)$ is smaller than $1$ at the dip near
$\Delta_{a}/\omega_{b}=0$ in the blue solid curve of
Fig.~\ref{fig3}(a), so the photon blockade can be observed. If the
value of $g$ is much smaller than the transition frequency
$\omega_{q}$, the TLS has a small effect on the photon blockade in
the blue solid curve of Fig.~\ref{fig3}(a) [compare with the black
dashed curve in Fig.~\ref{fig5}(c)]. If the value of $g$ becomes
larger, the TLS leads to two new dips (photon blockade) and
several peaks (photon tunneling) in the green solid curve of
Fig.~\ref{fig3}(b). This results from the counter-terms
$(b\sigma^{+}+\sigma_{-}b^{\dagger})$ which can be understood by
comparing the green solid and brown dashed curves of
Fig.~\ref{fig3}(b). When the value of coupling strength $g$ is
larger than the transition frequency of the TLS, that is
$g>\omega_{q}$, more dips and peaks appear in the red solid curve
of Fig.~\ref{fig3}(c). And the minimum value of $g^{(2)}(0)$ near
$\Delta_{a}/\omega_{b}=0$ is larger than $1$, so the photon
blockade in this regime vanishes in ultra-strong coupling regime.
Actually, a similar phenomenon of photon blockade in the
ultrastrong coupling regime has been studied in circuit
QED~\cite{Hartmann}.

\begin{figure}
\includegraphics[bb=20 195 560 650, width=8.5 cm, clip]{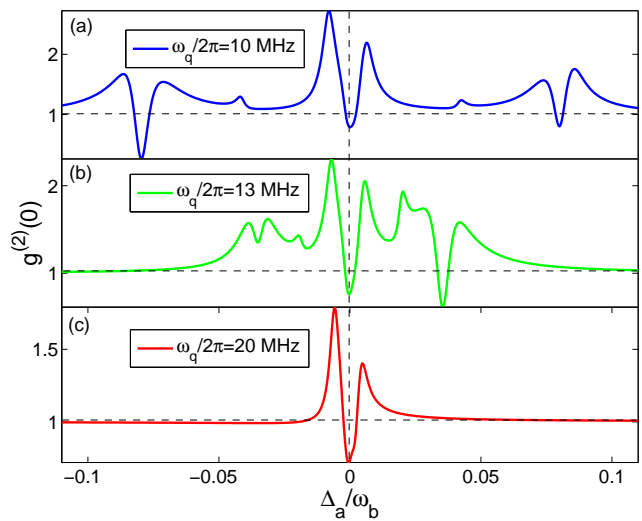}
\caption{(Color online) Equal-time second-order degree of
coherence $g^{(2)}(0)$ as a function of
$\Delta_{a}/\omega_{b}=(\omega_{a}-\omega_{d})/\omega_{b}$ in the steady-state limit.
 The curves in the three panels correspond to different transition
frequencies of the TLS: (a) $\omega_{q}/(2\pi)=10$ MHz; (b)
$\omega_{q}/(2\pi)=13$ MHz; and (c) $\omega_{q}/(2\pi)=20$ MHz.
The other parameters are as in Fig.~\ref{fig3}, except
$g/(2\pi)=4$ MHz.}\label{fig4}
\end{figure}

Figure~\ref{fig4} describes the effect of the TLS transition
frequency on the photon blockade in the hybrid system. The blue
solid curve of Fig.~\ref{fig4}(a) describes $g^{(2)}(0)$ when the
mechanical resonator interacts resonantly with a TLS, while the
green solid [in Figs.~\ref{fig4}(b)] and red solid curves [in
Fig.~\ref{fig4}(c)] correspond to the detuning cases. When the
mechanical mode and the TLS are in the detuning regime, the
positions of the left and right dips (relative to the point
$\Delta_{a}/\omega_{b}=0$) are changed in the green solid curve of
Fig.~\ref{fig4}(b). The minimum value of $g^{(2)}(0)$ of the left
dip  becomes larger than $1$, so the photon blockade disappears
near this point. But the photon blockade near the right dip is
enhanced. If the detuning $|\omega_{q}-\omega_{b}|$ is larger than
the coupling strength $g$, all the dips and peaks induced by the
TLS disappear in the red solid curve of Fig.~\ref{fig4}(c), in
this case the photon blockade is similar to that of standard
optomechanical systems [see the black solid curve in
Fig.~\ref{fig5}(c)].

The effect of the decay rate $\gamma_{q}$ on the photon blockade
of optomechanical systems is discussed in Fig.~\ref{fig5}. If the
value of $\gamma_{q}$ becomes larger, the dip near
$\Delta_{a}/\omega_{b}=0$ is almost invariant, but the left and
right dips (relative to the point $\Delta_{a}/\omega_{b}=0$)
change greatly. The minimum value of  $g^{(2)}(0)$ near the right
dip is even larger than $1$, so the photon blockade disappears in
this regime. The photon blockade near the left dip will also
vanish if the value of $\gamma_{q}$ continues to increase. If the
decay rate becomes very large, all the new dips and peaks induced
by the TLS vanish, and the photon blockade in the red solid curve
of Fig.~\ref{fig5}(c) is then almost then same to that of standard
optomechanical system [see the black dashed curve in
Fig.~\ref{fig5}(c)].

\begin{figure}
\includegraphics[bb=10 195 565 650, width=8.6 cm, clip]{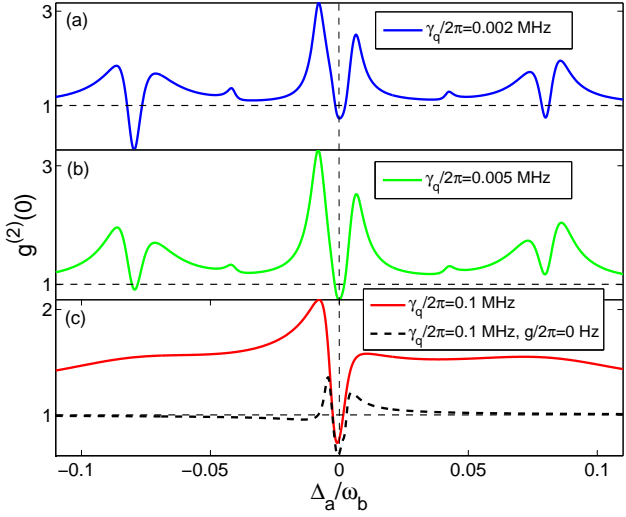}
\caption{(Color online) Equal-time second-order degree of
coherence $g^{(2)}(0)$ as a function of
$\Delta_{a}/\omega_{b}=(\omega_{a}-\omega_{d})/\omega_{b}$ in the steady-state limit.
 The curves in three panels correspond to different decay rates of TLS:
(a) $\gamma_{q}/(2\pi)=0.002$ MHz; (b) $\gamma_{q}/(2\pi)=0.005$
MHz; and (c) $\gamma_{q}/(2\pi)=0.1$ MHz. Here $g/(2\pi)=4 $ MHz
for the solid curves, while $g/(2\pi)=0 $ Hz for black dashed
curve in panel (c), and the other parameters are same as in
Fig.~\ref{fig3}. }\label{fig5}
\end{figure}

\begin{figure}
\includegraphics[bb=30 205 530 635, width=7.25 cm, clip]{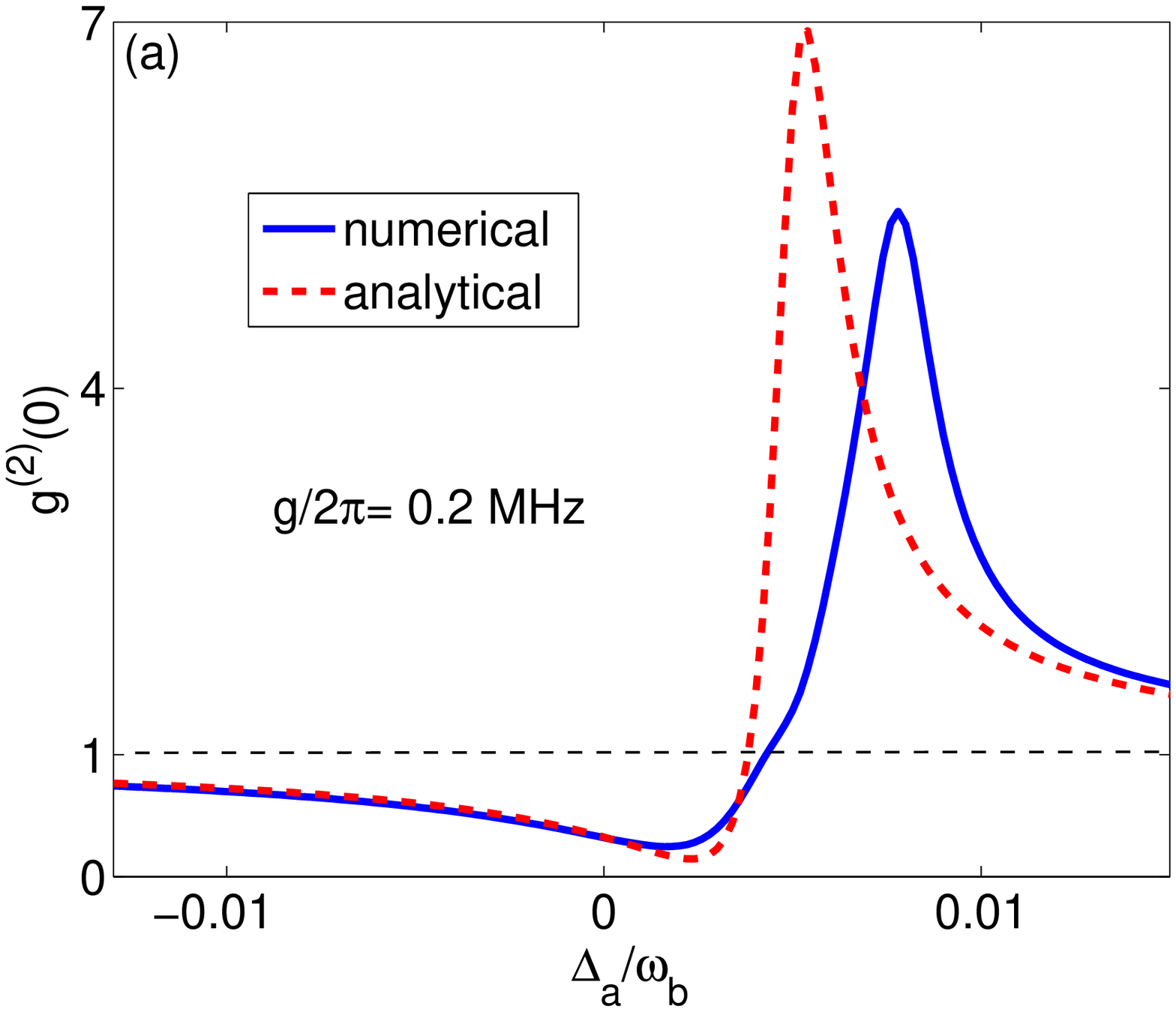}\\
\includegraphics[bb=30 205 530 620, width=7.4 cm, clip]{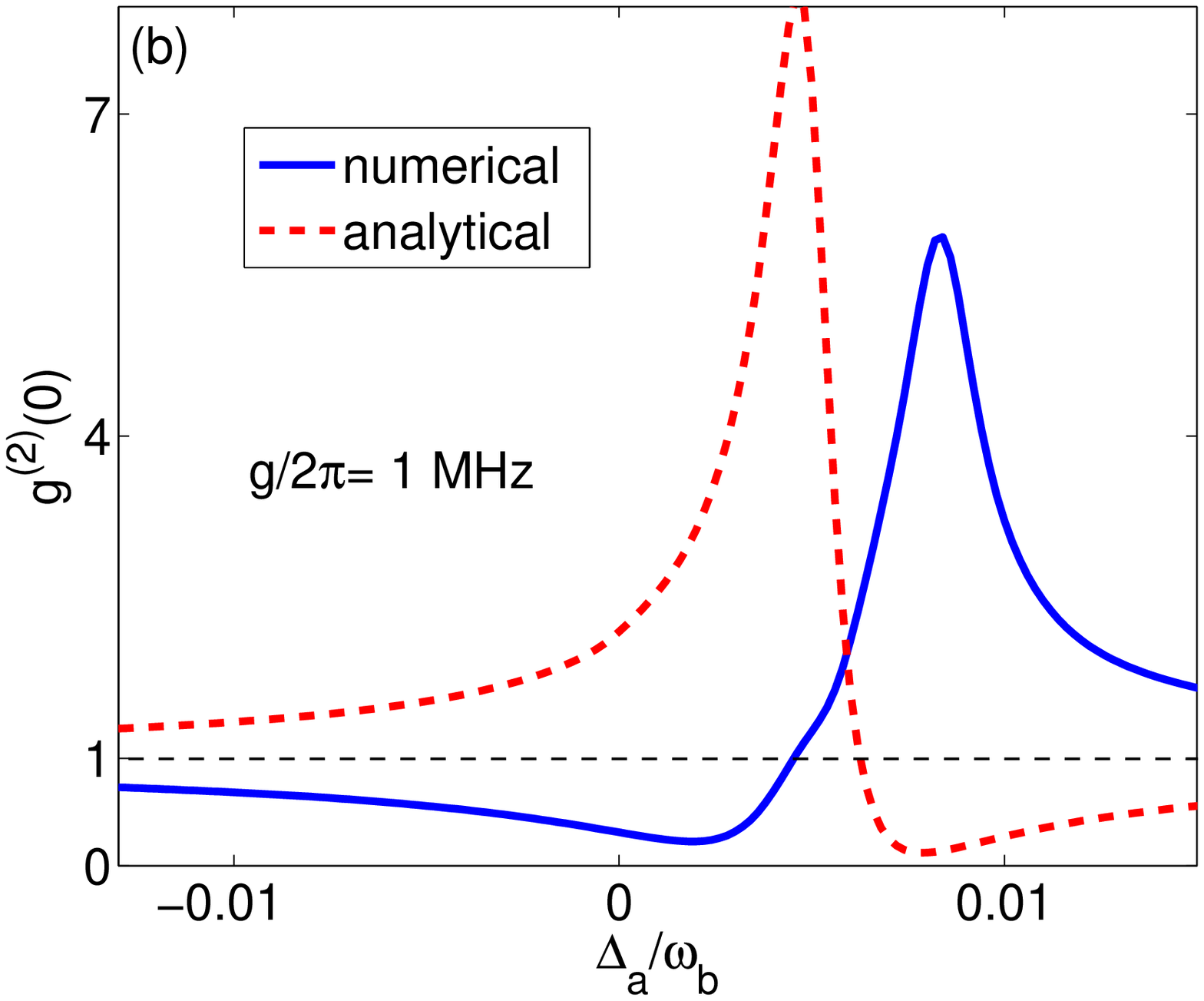}
\caption{(Color online) Equal-time second-order degree of
coherence $g^{(2)}(0)$ as a function of $\Delta_{a}/\omega_{b}$ at
zero temperature in the steady-state limit. The blue solid curves correspond to our
numerical precise solutions of the master equation, while the red
dashed curves are plotted with the analytical result in
Eq.~(\ref{eq:9}). Both figures are plotted by substituting
$\sigma_{z}$ with $\sigma_{+}\sigma_{-}$ in Eq.~(\ref{eq:3}). Here
$\chi/(2\pi)=0.5$ MHz, $|\Omega|/(2\pi)=0.01$ MHz, and $T=0$ K,
while  other parameters are the same as in Fig.~\ref{fig3}.
}\label{fig6}
\end{figure}

In Fig.~\ref{fig6}, we compare the results obtained numerically by
the master equation in Eq.~(\ref{eq:4}) with those obtained
analytically in Eq.~(\ref{eq:9}). We plot $g^{(2)}(0)$ as a
function of the detuning $\Delta_{a}/\omega_{b}$ for $T=0$ K. The
blue solid curves in Figs.~\ref{fig6}(a) and~\ref{fig6}(b) are
plotted using the master equation, while the red dashed curves are
plotted with the analytical result in Eq.~(\ref{eq:9}). The
results of two methods are almost the same for Fig.~\ref{fig6}(a).
If the coupling strength $g$ becomes larger, the deviations
between the blue solid and red dashed curves in Fig.~\ref{fig6}(b)
becomes larger. This difference originates from the approximation
when Eq.~(\ref{eq:9}) was derived, because, contrary to our precise numerical calculations, we have assumed
in our analytical approach that (i) the system evolution is pure and (ii) some transition
processes, such as $|0,2,\downarrow\rangle$,
$|0,2,\uparrow\rangle$, $|1,2,\downarrow\rangle$, etc., were
neglected.

Therefore, we conclude that the coupling strength (to the
mechanical mode), transition frequency, and the decay rate of the
TLS can be used to tune the photon blockade and tunneling of
optomechanical systems.

\section{Conclusions}\label{conclusion}

We have studied single-photon blockade and tunneling(corresponding
to multi-photon blockade) of a hybrid system consisting of an
optomechanical cavity and a TLS. We find that the photon blockade
of the optomechanical device is significantly affected by a TLS
when it is coupled to the mechanical resonator. Compared with the
results of only the optomechanical cavity, the TLS shifts and
splits the peaks and dips of the second-order degree of coherence
of the cavity field in the optomechanical subset. We also find
that the TLS gives rise to several new peaks and dips in the
second-order degree of coherence of the cavity field in the hybrid
system.

If the coupling strength (between the mechanical mode and the TLS)
is comparable or larger than the transition frequency of the TLS,
new blockade dips and resonant peaks appear for  the second degree
of coherence. Moreover the  new blockade dips and resonant peaks
can be tuned if we change the transition frequency or the decay
rate of the TLS. The photon anti-bunching of hybrid systems can
also be tuned if we change the parameters of the TLS. That is, our
study may provide a new method to control and tune the
nonlinearity and nonclassical effect of the cavity field of the
optomechanical system by coupling to a tunable TLS. Our
calculation may also provide an approach to detect a low-frequency TLS
(which might be a defect in a mechanical resonator)
 using optomechanics.

\section{Acknowledgement}

We thank Anton F. Kockum for valuable suggestions to the
manuscript. Hui Wang thanks Nan Yang and Jing Zhang for technical
support. Y.X.L. is supported by the National Natural Science
Foundation of China under Grant No. 61328502, the National Basic
Research Program of China (973 Program) under Grant No.
2014CB921401, the Tsinghua University Initiative Scientific
Research Program, and the Tsinghua National Laboratory for
Information Science and Technology (TNList) Cross-discipline
Foundation. A.M. acknowledges a long-term fellowship from the
Japan Society for the Promotion of Science (JSPS). A.M. is
supported by the Polish National Science Centre under the grants
No. DEC-2011/02/A/ST2/00305 and No. DEC-2011/03/B/ST2/01903. F.N.
is partially supported by the RIKEN iTHES Project, the MURI Center
for Dynamic Magneto-Optics via the AFOSR award number
FA9550-14-1-0040, the IMPACT program of JST, and a Grant-in-Aid
for Scientific Research (A).

\appendix*
\section{The expansion coefficients of Eq.~(\ref{eq:7})}\label{App-A}

In the weak pumping limit, only low excited states of photons and
phonons are occupied. If the temperature is extremely low, the
quantum state of the hybrid system can be written as a sum  of the
finite orthogonal basis states given in Eq.~(\ref{eq:5}).
Considering the effect of environment noises, the effective
non-Hermitian Hamiltonian of the hybrid system can be obtained in
Eq.~(\ref{eq:7}). In the steady state case, we can set $
d|\psi\rangle/dt=0$. Substituting Eqs.~(\ref{eq:5})
and~(\ref{eq:7}) into Eq.~(\ref{eq:6}), we obtain linear equations
about the expanding coefficients of Eq.~(\ref{eq:5}) as follow:
\begin{eqnarray}
0&=&\left(\omega^{\prime}_{q}/2\right)C_{00\uparrow}+g C_{01\downarrow}-i\Omega^{\ast} C_{10\uparrow}\label{eq:A1},\\
0&=&\Delta^{-}_{1}C_{01\downarrow}+ g C_{00\uparrow}-i\Omega^{\ast} C_{11\downarrow}\label{eq:A2},\\
0&=&\Delta^{+}_{1} C_{01\uparrow}+ g C_{00\downarrow}-i\Omega^{\ast} C_{11\uparrow}\label{eq:A3},\\
0&=&\Delta^{-}_{2}C_{10\downarrow}-\chi C_{11\downarrow}+g C_{11\uparrow}+i\Omega C_{00\downarrow}\nonumber\\
& &-i\sqrt{2}\Omega^{\ast} C_{20\downarrow}\label{eq:A4},\\
0&=&\Delta^{+}_{2}C_{10\uparrow}-\chi C_{11\uparrow}+g C_{11\downarrow}+i\Omega C_{00\uparrow}\nonumber\\
& &-i\sqrt{2}\Omega^{\ast} C_{20\uparrow},\label{eq:A5},\\
0&=&\Delta^{-}_{3}C_{11\downarrow}-\chi C_{10\downarrow}+ g C_{10\uparrow}+i\Omega C_{01\downarrow}\nonumber\\
& &-i\sqrt{2}\Omega^{\ast} C_{21\downarrow},\label{eq:A6},\\
0&=&\Delta^{+}_{3}C_{11\uparrow}-\chi C_{10\uparrow}+g C_{10\downarrow}+i\Omega C_{01\uparrow},\label{eq:A7}\\
0&=&\Delta^{-}_{4}C_{20\downarrow}-2\chi C_{21\downarrow}+i\sqrt{2}\Omega C_{10\downarrow},\label{eq:A8}\\
0&=&\Delta^{+}_{4}C_{20\uparrow}+ g C_{21\downarrow}+i\sqrt{2}\Omega C_{10\uparrow},\label{eq:A9}\\
0&=&\Delta_{5}C_{21\downarrow}-2\chi C_{20\downarrow}+ g
C_{20\uparrow}+i\sqrt{2}\Omega C_{11\downarrow},\label{eq:A10}
\end{eqnarray}
with the definitions of $\Delta_{i} (i=1,2,\ldots,9)$ being
\begin{eqnarray}\label{eq:A11}
\Delta^{\mp}_{1}&=&\omega^{\prime}_{m}\mp\omega^{\prime}_{q}/2,\nonumber\\
\Delta^{\mp}_{2}&=&\Delta^{\prime}_{a}\mp\omega^{\prime}_{q}/2,\nonumber\\
\Delta^{\mp}_{3}&=&\Delta^{\prime}_{a}+\omega^{\prime}_{m}\mp\omega^{\prime}_{q}/2,\nonumber\\
\Delta^{\mp}_{4}&=&2\Delta^{\prime}_{a}\mp\omega^{\prime}_{q}/2,\nonumber\\
\Delta_{5}&=&2\Delta^{\prime}_{a}+\omega^{\prime}_{m}-\omega^{\prime}_{q}/2.
\end{eqnarray}
The equation
$0=-\omega^{\prime}_{q}C_{00\downarrow}/2-i\Omega^{\ast}
C_{10\downarrow}$ has no physical meaning if
$|\Omega|\longrightarrow 0$, so it can be neglected. The system
has a probability to remain in the ground state
$|0,0,\downarrow\rangle$ in the weak-pumping limit, so we can set
$C_{00\downarrow}=1$ (then the expansion coefficients are
unnormalized). The terms $(i\sqrt{2}\Omega^{\ast}
C_{20\downarrow})$ in Eq.~(\ref{eq:A4}), $(i\Omega^{\ast}
C_{20\uparrow})$ in Eq.~(\ref{eq:A5}), and
$(-i\sqrt{2}\Omega^{\ast} C_{21\downarrow})$ in Eq.~(\ref{eq:A6})
are of higher-order in $\Omega$, so they can be neglected.

Through  some calculations, we can obtain the solutions of the
expansion coefficients in Eq.~(\ref{eq:5}). The corresponding
coefficients $\eta_{i} (i=1,2,\ldots,7)$ are given by
\begin{eqnarray}\label{eq:A12}
\eta_{1}&=&\left(\chi\Delta^{+}_{1}\lambda_{11}+g\Delta^{+}_{1}\lambda_{9}-\Delta^{-}_{2}g\lambda_{9}\right)/D_{1},\nonumber\\
\eta_{2}&=&-\left(g\Delta^{+}_{1}\lambda_{8}+g\Delta^{-}_{2}\lambda_{8}+\chi\Delta^{+}_{1}\lambda_{10}\right)/D_{1},\nonumber\\
\eta_{3}&=&\left(\chi \eta_{1}-g \eta_{2}-1\right)/\Delta^{-}_{2},\nonumber\\
\eta_{4}&=&\left(\chi \eta_{2}-\lambda_{4}\eta_{1}\right)/\lambda_{3},\nonumber\\
\eta_{5}&=&\left(\lambda_{13}\lambda_{14}+2\chi g\lambda_{15}\right)/D_{2},\nonumber\\
\eta_{6}&=&-\left(\lambda_{12}\lambda_{15}+2\chi g\lambda_{14}\right)/D_{2},\nonumber\\
\eta_{7}&=&2\chi\left(\lambda_{13}\lambda_{14}+2\chi g\lambda_{15}\right)/(\Delta_{5}D_{2})+\sqrt{2}\eta_{1}/\Delta_{5}\nonumber\\
& &+g\left(\lambda_{12}\lambda_{15}+2\chi
g\lambda_{14}\right)/(\Delta_{5}D_{2}),
\end{eqnarray}
with
$D_{1}=\Delta^{+}_{1}\Delta^{-}_{2}\left(\lambda_{9}\lambda_{10}-\lambda_{8}\lambda_{11}\right)$
and $D_{2}=\lambda_{12}\lambda_{13}-4\chi^{2} g^{2}$. The
parameters $\lambda_{i}(i=1,2,\ldots,15)$ in the above equations
are
\begin{eqnarray}\label{eq:A13}
\lambda_{1}&=&\Delta^{-}_{1}-2g^{2}/\omega^{\prime}_{q},\nonumber\\
\lambda_{2}&=&\Delta^{+}_{2}-2|\Omega|^{2}/\omega^{\prime}_{q},\nonumber\\
\lambda_{3}&=&\lambda_{2}-4|\Omega|^{2}g^{2}/\left(\lambda_{1}\omega^{\prime2}_{q}\right),\nonumber\\
\lambda_{4}&=&g +2g|\Omega|^{2}/\left(\lambda_{1}\omega^{\prime}_{q}\right),\nonumber\\
\lambda_{5}&=&\Delta^{-}_{3}-|\Omega|^{2}/\lambda_{1},\nonumber\\
\lambda_{6}&=&g +2|\Omega|^{2}g/\lambda_{1}\omega^{\prime}_{q},\nonumber\\
\lambda_{7}&=&\Delta^{+}_{3}- |\Omega|^{2}/\Delta^{+}_{1},\nonumber\\
\lambda_{8}&=&\lambda_{5}-\chi^{2}/\Delta^{-}_{2}-\lambda_{4}\lambda_{6}/\lambda_{3},\nonumber\\
\lambda_{9}&=&g\chi/\Delta^{-}_{2} +\chi\lambda_{6}/\lambda_{3},\nonumber\\
\lambda_{10}&=&g\chi/\Delta^{-}_{2}+\chi\lambda_{4}/\lambda_{3},\nonumber\\
\lambda_{11}&=&\lambda_{7}-\chi^{2}/\lambda_{3} -g^{2}/\Delta^{-}_{2},\nonumber\\
\lambda_{12}&=&\Delta^{-}_{4}\Delta_{5}-4\chi^{2},\nonumber\\
\lambda_{13}&=&\Delta^{+}_{4}\Delta_{5}- g^{2},\nonumber\\
\lambda_{14}&=&\sqrt{2}(2\chi\eta_{1}+\Delta_{5}\eta_{3}),\nonumber\\
\lambda_{15}&=&\sqrt{2}g\eta_{1}-\sqrt{2}\Delta_{5} \eta_{4}.
\end{eqnarray}
With the expressions of $\eta_{i} (i=1,2,\ldots,7)$ in
Eqs.~(\ref{eq:A12}), we can obtain the analytical result of the
second-order degree of coherence in Eq.~(\ref{eq:9}) in the
weak-pumping and low-temperature limit.



\begin{thebibliography}{99}

\bibitem{Marquardt}
M. Aspelmeyer, T. J. Kippenberg, and F. Marquardt, \Title{Cavity
optomechanics,} Rev. Mod. Phys.~\textbf{86}, 1391 (2014).

\bibitem{Aspelmeyer}
M. Aspelmeyer, P. Meystre, and   K. Schwab, \Title{Quantum
optomechanics,} Physics Today~\textbf{65}, 29 (2012).

\bibitem{PR1}
M. P. Blencowe, \Title{Quantum electromechanical systems,} Phys.
Rep.~\textbf{395}, 159 (2004).

\bibitem{PR2}
M. Poot and H. S. J. van der Zant, \Title{Mechanical systems in
the quantum regime,} Phys. Rep.~\textbf{511}, 273 (2012).

\bibitem{review1}
T. J. Kippenberg and K. J. Vahala, \Title{Cavity optomechanics:
back-action at the mesoscale,} Science~\textbf{321}, 1172 (2008).

\bibitem{Schwab1}
K. C. Schwab and M. L. Roukes, \Title{Putting Mechanics into
Quantum Mechanics,} Phys. Today~\textbf{58}, 36 (2005).

\bibitem{Schwab2}
A. Naik, O. Buu, M. D. LaHaye,  A. D. Armour, A. A. Clerk, M. P.
Blencowe, and  K. C. Schwab, \Title{Cooling a nanomechanical
resonator with quantum back-action,}  Nature
(London)~\textbf{443}, 193 (2006).

\bibitem{Painter}
J. Chan, T. P. M. Alegre, A. H. Safavi-Naeini, J. T. Hill, A.
Krause,  S. Gr\"{o}blacher, M. Aspelmeyer, and O. Painter,
\Title{Laser cooling of a nanomechanical oscillator into its
quantum ground state,} Nature (London)~\textbf{478}, 89 (2011).

\bibitem{Meystre}
P. Meystre, \Title{Cool vibrations,} Science~\textbf{333}, 832
(2011).

\bibitem{Tang1}
M. Bagheri, M. Poot, L. Fan, F. Marquardt, and H. X. Tang,
\Title{Photonic Cavity Synchronization of Nanomechanical
Oscillators,} Phys. Rev. Lett.~\textbf{111}, 213902 (2013).

\bibitem{Tang2}
X. Sun, J. Zheng, M. Poot, C. W. Wong, and H. X. Tang,
\Title{Femtogram Doubly Clamped Nanomechanical Resonators Embedded
in a High-Q Two-Dimensional Photonic Crystal Nanocavity,} Nano
Lett.~\textbf{12}, 2299 (2012).

\bibitem{NatureMilestones}
P. Rodgers, \Title{ Cavity optomechanics: Mirror finish}, Nature Milestones~\textbf{9}, S20 (2010).
\Title{doi:10.1038/nmat2660}

\bibitem{Xin}
X. Y. L\"u, Y. Wu, J. R. Johansson, H. Jing, J.  Zhang, and F.
Nori, \Title{Squeezed Optomechanics with Phase-matched
Amplification and Dissipation}, Phys. Rev. Lett. ~\textbf{114},
093602 (2015).

\bibitem{Braginsky92}
V.~B.~Braginsky and F.~Ya.~Khalili, \textit{Quantum Measurements}
(Cambridge University Press, Cambridge, England, 1992).

\bibitem{Caves}
C.~M.~Caves, K.~S.~Thorne, R.~W.~P.~Drever, V.~D.~Sandberg, and
M.~Zimmermann, \Title{On the measurement of a weak classical force
coupled to a quantum-mechanical oscillator. ,} Rev. Mod.
Phys.~\textbf{52}, 341 (1980).

\bibitem{Bocko96}
M.~F.~Bocko and R.~Onofrio, \Title{On the measurement of a weak
classical force coupled to a harmonic oscillator: experimental
progress,} Rev. Mod. Phys.~\textbf{68}, 755 (1996).

\bibitem{Buks06}
E.~Buks and B.~Yurke, \Title{Mass detection with a nonlinear
nanomechanical resonator,} Phys. Rev. E~\textbf{74}, 046619
(2006).

\bibitem{OConnell10}
A. D. O'Connell \etal, \Title{Quantum ground state and
single-phonon control of a mechanical resonator,} Nature
(London)~\textbf{464}, 697 (2010).

\bibitem{Lin11}
L. Tian and H. Wang, \Title{Optical wavelength conversion of
quantum states with optomechanics,} \pra~\textbf{82}, 053806
(2010).

\bibitem{Hill}
J. T. Hill, A. H. Safavi-Naeini, J. Chan, and O. Painter,
\Title{Coherent optical wavelength conversion via cavity
optomechanics,} Nat. Comm.~\textbf{3}, 1196 (2012).

\bibitem{Bochmann}
J. Bochmann, A. Vainsencher, D. D. Awschalom  and A. N. Cleland,
\Title{Nanomechanical coupling between microwave and optical
photons,} Nat. Phys.~\textbf{9}, 712 (2013).

\bibitem{Lin12}
L. Tian, \Title{Adiabatic state conversion and pulse transmission
in optomechanical systems,} Phys. Rev. Lett.~\textbf{108}, 153604
(2012).

\bibitem{Lin13}
L. Tian, \Title{Optoelectromechanical transducer: Reversible
conversion between microwave and optical photons,} Ann. Phys.
(Berlin)~\textbf{527}, 1 (2014).

\bibitem{xu-phonon1}
X. W. Xu, H. Wang, J. Zhang, and Y. X. Liu, \Title{Engineering of
nonclassical motional states in optomechanical systems,}
\pra~\textbf{88}, 063819 (2013).

\bibitem{xu-phonon2}
X. W. Xu, Y. J. Zhao, and Y. X. Liu, \Title{Entangled-state
engineering of vibrational modes in a multimembrane optomechanical
system,} \pra~\textbf{88}, 022325 (2013).

\bibitem{Long}
F. C. Lei, M. Gao, C. G. Du, S. Y. Hou, X. Yang, and G. L. Long,
\Title{Engineering optomechanical normal modes for single-phonon
transfer and entanglement preparation}, J.  Opt. Soc. Am.
B~\textbf{32}, 588 (2015).

\bibitem{Agarwal}
S. Huang and G. S. Agarwal, \Title{Normal-mode splitting in a
coupled system of a nanomechanical oscillator and a parametric
amplifier cavity,} \pra~\textbf{80} 033807 (2009).

\bibitem{Weis} S. Weis, R. Rivi\`ere, S. Del\'eglise, E. Gavartin,
O. Arcizet, A. Schliesser,  and T. J. Kippenberg,
\Title{Optomechanically Induced Transparency,}
Science~\textbf{330}, 1520 (2010).

\bibitem{Liao1} J. Q. Liao, H. K. Cheung, and C. K. Law,
\Title{Spectrum of single-photon emission and scattering in cavity
optomechanics,}  \pra~\textbf{85}, 025803 (2012).

\bibitem{Ma} P.C. Ma, Jian-Qi Zhang, Yin Xiao, Mang Feng, and Zhi-Ming Zhang,
\Title{Tunable double optomechanically induced transparency in an
optomechanical system}, Phys. Rev. {\bf A 90}, 043825 (2014).

\bibitem{Qi}
Q. Wu, J. Q. Zhang, J. H. Wu, M. Feng, Z. M. Zhang, \Title{Tunable
multi-channel inverse optomechanically induced transparency},
arXiv:1504.05359.

\bibitem{Akram3}
M.J. Akram,  K. Naseer, and F. Saif, \Title{Effcient tunable
switch from slow light to fast light in quantum
opto-electromechanical system}, arXiv:1503.01951.

\bibitem{Pirkkalainen}
J.-M. Pirkkalainen, S. U. Cho, J. Li, G. S. Paraoanu, P. J.
Hakonen, and M. A. Sillanp\"{a}\"{a}, \Title{Hybrid circuit cavity
quantum electrodynamics with a micromechanical resonator,} Nature
(London)~\textbf{494}, 211(2013).

\bibitem{Ciuti}
J. Restrepo, C. Ciuti, and I. Favero, \Title{Single-Polariton
Optomechanics,}  \prl~\textbf{112}, 013601 (2014).

\bibitem{Jia}
W. Z. Jia and Z. D. Wang, \Title{Single-photon transport in a
one-dimensional waveguide coupling to a hybrid atom-optomechanical
system,} \pra~\textbf{88}, 063821 (2013).

\bibitem{Akram1}
M. J. Akram, M. M. Khan, and F. Saif, \Title{Tunable fast and slow light in a
hybrid optomechanical system,} \pra~\textbf{92}, 023846 (2015).

\bibitem{Akram2}
M. J. Akram, F. Ghafoor, and F. Saif, \Title{Electromagnetically
induced transparency and tunable {F}ano resonances in hybrid
optomechanics,} J. Phys. B: At. Mol. Opt. Phys. {\bf 48}, 065502
(2015).

\bibitem{hui}
H. Wang, H. C. Sun, J. Zhang, and Y. X. Liu, \Title{Transparency
and amplification in a hybrid system of the mechanical resonator
and circuit QED,} Sci. China-Phys. Mech. Astron.~\textbf{55}, 2264
(2012).

\bibitem{ZLXiang}
Z. L. Xiang, S. Ashhab, J. Q. You, and F. Nori, \Title{Hybrid
quantum circuits: Superconducting circuits interacting with other
quantum systems,} Rev. Mod. Phys.~\textbf{85}, 623 (2013).

\bibitem{Tdefects}
L. Tian, \Title{Cavity cooling of a mechanical resonator in the
presence of a two-level-system defect,} \prb~\textbf{84}, 035417
(2011).

\bibitem{hybrid2}
T. Ramos, V. Sudhir, K. Stannigel, P. Zoller, and T. J.
Kippenberg, \Title{Nonlinear quantum optomechanics via individual
intrinsic two-level defects,} \prl~\textbf{110}, 193602 (2013).

\bibitem{Wang}
H. Wang, X. Gu, Y. X. Liu, A. Miranowicz, and F. Nori,
\Title{Optomechanical analog of two-color
electromagnetically-induced transparency: {P}hoton transmission
through an optomechanical device with a two-level system}
\pra~\textbf{90}, 023817 (2014).

\bibitem{Tian92}
L. Tian and H. J. Carmichael, \Title{Quantum trajectory
simulations of two-state behavior in an optical cavity containing
one atom,} \pra \textbf{46}, R6801 (1992).

\bibitem{Leonski94}
W. Leo\'nski and R. Tana\'s, \Title{Possibility of producing the
one-photon state in a kicked cavity with a nonlinear Kerr medium,}
\pra \textbf{49}, R20 (1994).

\bibitem{Miran96}
A. Miranowicz, W. Leo\'nski, S. Dyrting, and R. Tana\'s,
\Title{Quantum state engineering in finite-dimensional Hilbert
space,} Acta Phys. Slov.  \textbf{46}, 451 (1996).

\bibitem{Imamoglu97}
A.~Imamo\={g}lu, H.~Schmidt, G. Woods, and M. Deutsch,
\Title{Strongly interacting photons in a nonlinear cavity,} \prl
\textbf{79}, 1467 (1997).

\bibitem{Rebic99}
S. Rebi\'{c}, S. M. Tan, A. S. Parkins, and D. F. Walls,
\Title{Large Kerr nonlinearity with a single atom,} J. Opt. B
\textbf{1}, 490 (1999).

\bibitem{Kim99}
J. Kim, O. Bensen, H. Kan, and Y. Yamamoto, \Title{A single-photon
turnstile device,} Nature (London) \textbf{397}, 500 (1999).

\bibitem{Rebic02}
S. Rebi\'c, A. S. Parkins, and S. M. Tan, \Title{Photon statistics
of a single-atom intracavity system involving electromagnetically
induced transparency,}  \pra \textbf{65}, 063804 (2002).

\bibitem{Smolyaninov02}
I. I. Smolyaninov, A. V. Zayats, A. Gungor, and C. C. Davis,
\Title{Single-photon tunneling via localized surface plasmons,}
\prl \textbf{88}, 187402 (2002).

\bibitem{Birnbaum05}
K. M. Birnbaum, A. Boca, R. Miller, A. D. Boozer, T. E. Northup,
and H. J. Kimble, \Title{Photon blockade in an optical cavity with
one trapped atom,} Nature (London) \textbf{436}, 87 (2005).


\bibitem{Hoffman11}
A. J. Hoffman, S. J. Srinivasan, S. Schmidt, L. Spietz, J.
Aumentado, H. E. Tureci, and A. A. Houck, \Title{Dispersive photon
blockade in a superconducting circuit,} \prl \textbf{107}, 053602
(2011).

\bibitem{Lang11}
C. Lang \etal, \Title{Observation of resonant photon blockade at
microwave frequencies using correlation function measurements,}
\prl \textbf{106}, 243601 (2011).

\bibitem{Didier11}
N. Didier, S. Pugnetti, Y. M. Blanter, and R. Fazio,
\Title{Detecting phonon blockade with photons,} \prb \textbf{84},
054503 (2011).

\bibitem{Liu14}
Y. X. Liu, X. W. Xu, A. Miranowicz, and F. Nori, \Title{From
blockade to transparency: controllable photon transmission through
a circuit QED system,} Phys. Rev. A~\textbf{89}, 043818 (2014).

\bibitem{Rabl}
P. Rabl, \Title{Photon blockade effect in optomechanical systems,}
\prl \textbf{107}, 063601 (2011).

\bibitem{Nunnenkamp11}
A. Nunnenkamp, K. B\o{}rkje, and S. M. Girvin,
\Title{Single-photon optomechanics,} \prl \textbf{107}, 063602
(2011).

\bibitem{Liao13}
J.Q. Liao and F. Nori, \Title{Photon blockade in quadratically
coupled optomechanical systems,} \pra \textbf{88}, 023853 (2013).

\bibitem{Adam13}
A. Miranowicz, M. Paprzycka, Y. X. Liu, J. Bajer, and  F. Nori,
\Title{Two-photon and three-photon blockades in driven nonlinear
systems,} \pra~\textbf{87}, 023809 (2013).

\bibitem{Adam14a} A. Miranowicz, M. Paprzycka, A. Pathak, and F. Nori,
\Title{Phase-space interference of states optically truncated by
quantum scissors}, \pra~\textbf{89}, 033812 (2014).

\bibitem{Adam14b}
A. Miranowicz, J. Bajer, M. Paprzycka, Y. X. Liu, A. M. Zagoskin,
and F. Nori, \Title{State-dependent photon blockade via
quantum-reservoir engineering,} \pra~\textbf{90}, 033831 (2014).

\bibitem{Hovsepyan14}
G. H. Hovsepyan, A. R. Shahinyan, and G. Y. Kryuchkyan,
\Title{Multiphoton blockades in pulsed regimes beyond stationary
limits,} \pra~\textbf{90}, 013839 (2014).

\bibitem{Faraon08}
A. Faraon, I. Fushman, D. Englund, N. Stoltz, P. Petroff, and J.
Vu\v{c}kovi\'c, \Title{Coherent generation of non-classical light
on a chip via photon-induced tunnelling and blockade,} Nat. Phys.
{\bf 4}, 859 (2008).

\bibitem{Majumdar12}
A. Majumdar, M. Bajcsy, and J. Vu\v{c}kovi\'c, \Title{Probing the
ladder of dressed states and nonclassical light generation in
quantum-dot-cavity QED,} \pra \textbf{85}, 041801(R) (2012).

\bibitem{Schuster08}
I. Schuster, A. Kubanek, A. Fuhrmanek, T. Puppe, P. W. H. Pinkse,
K. Murr, and G. Rempe, \Title{Nonlinear spectroscopy of photons
bound to one atom,} Nature Phys. \textbf{4}, 382 (2008).

\bibitem{Kubanek08}
A. Kubanek, A. Ourjoumtsev, I. Schuster, M. Koch, P. W. H. Pinkse,
K. Murr, and G. Rempe, \Title{Two-Photon Gateway In One-Atom
Cavity Quantum Electrodynamics,} \prl \textbf{101}, 203602 (2008).

\bibitem{alsing1}
P. Alsing, D.-S. Guo, and H. J. Carmichael, \Title{Dynamic Stark
effect for the Jaynes-Cummings system,} \pra~\textbf{45}, 5135
(1992).

\bibitem{Kim}
M. S. Kim, F. A. M. De Oliveira, and P. L. Knight, \Title{The
Interaction of Displaced Number State and Squeezed Number State
Fields with Two-level Atoms,} J. Mod. Opt.~\textbf{37}, 659
(1990).

\bibitem{Cao}
X. Cao,  J. Q. You, H. Zheng, and F. Nori, \Title{A qubit strongly
coupled to a resonant cavity: asymmetry of the spontaneous
emission spectrum beyond the rotating wave approximation,} New J.
Phys.~\textbf{13}, 073002 (2011).

\bibitem{Zueco}
D. Zueco,  G. M. Reuther, S. Kohler, and P. H\"{a}nggi,
\Title{Qubit-oscillator dynamics in the dispersive regime:
Analytical theory beyond the rotating-wave approximation,}
\pra~\textbf{80}, 033846 (2009).

\bibitem{Niemczyk}
T. Niemczyk \etal, \Title{Circuit quantum electrodynamics in the
ultrastrong-coupling regime,}  Nat. Phys.~\textbf{6}, 772 (2010).

\bibitem{Crisp}
M. D. Crisp, \Title{Jaynes-Cummings model without the
rotating-wave approximation,} \pra~\textbf{43}, 2430(1991).

\bibitem{Tanas03}
R. Tana\'s, \Title{Nonclassical states of light propagating in
Kerr media,} in: \emph{Theory of Non-Classical States of Light}
ed. V. A. Dodonov and V. I. Man'ko (Taylor \& Francis, London,
2003) p. 267

\bibitem{HarocheBook}
S. Haroche and J. M. Raimond, \emph{Exploring the Quantum: Atoms,
Cavities and Photons} (Oxford University, Oxford, 2006).

\bibitem{Tanas83}
R. Tana\'s and S. Kielich, \Title{Self-squeezing of light
propagating through nonlinear optically isotropic media,} Opt.
Commun. \textbf{45}, 351 (1983);

\bibitem{Milburn86}
G. J. Milburn, \Title{Quantum and classical Liouville dynamics of
the anharmonic oscillator,} \pra~\textbf{33}, 674 (1986).

\bibitem{Yamamoto86}
Y. Yamamoto, N. Imoto, and S. Machida, \Title{Amplitude squeezing
in a semiconductor laser using quantum nondemolition measurement
and negative feedback,} \pra~\textbf{33}, 3243 (1986)

\bibitem{Tanas91}
R. Tana\'s, A. Miranowicz, and S. Kielich, \Title{Squeezing and
its graphical representations in the anharmonic oscillator model,}
Phys. Rev. A  {\bf 43}, 4014 (1991).

\bibitem{Yurke86}
B. Yurke and D. Stoler, \Title{Generating quantum mechanical
superpositions of macroscopically distinguishable states via
amplitude dispersion,} \prl~\textbf{57}, 13 (1986).

\bibitem{Tombesi87}
P. Tombesi and A. Mecozzi, \Title{Generation of macroscopically
distinguishable quantum states and detection by the
squeezed-vacuum technique,} \josab~\textbf{4}, 1700 (1987).

\bibitem{Miran90}
A. Miranowicz, R. Tana\'s, and S. Kielich, \Title{Generation of
discrete superpositions of coherent states in the anharmonic
oscillator model,} Quantum Opt.~\textbf{2}, 253 (1990); R.
Tana\'s, Ts. Gantsog, A. Miranowicz, and S. Kielich,
\Title{Quasi-probability distribution $Q(\alpha,\alpha^*)$ versus
phase distribution $P(\theta)$ in a description of superpositions
of coherent states,} \josab~\textbf{8}, 1576 (1991).

\bibitem{Kirchmair13}
G. Kirchmair, B. Vlastakis, Z. Leghtas, S. E. Nigg, H. Paik, E.
Ginossar, M. Mirrahimi, L. Frunzio, S. M. Girvin, and R. J.
Schoelkopf, \Title{Observation of quantum state collapse and
revival due to the single-photon Kerr effect,} Nature
(London)~\textbf{495}, 205 (2013).

\bibitem{Dodonov11}
A. V. Dodonov and V. V. Dodonov,  \Title{Strong modifications of
the field statistics in the cavity dynamical {C}asimir effect due
to the interaction with two-level atoms and detectors}, Phys.
Lett. A~\textbf{375}, 4261  (2011).

\bibitem{Dodonov10}
A. V. Dodonov,  \Title{Current status of the dynamical {C}asimir
effect}, Phys. Scr. \textbf{82}, 038105 (2010).

\bibitem{J1}
J. R. Johansson, G. Johansson, C. M. Wilson, and F. Nori,
\Title{Dynamical {C}asimir Effect in a Superconducting Coplanar
Waveguide}, \prl~\textbf{103}, 147003 (2009).

\bibitem{J2}
J. R. Johansson, G. Johansson, C. M. Wilson, and F. Nori,
\Title{Dynamical {C}asimir effect in superconducting microwave
circuits}, \pra~\textbf{82}, 052509 (2010).

\bibitem{Wilson11}
C. M. Wilson, G. Johansson, A. Pourkabirian, J. R. Johansson, T.
Duty, F. Nori, and P. Delsing, \Title{Observation of the dynamical
{C}asimir effect in a superconducting circuit,} Nature (London)
\textbf{479}, 376 (2011).

\bibitem{Dodonov12a}
A. V. Dodonov and V. V. Dodonov,  \Title{Dynamical {C}asimir
effect in a cavity with an $N$-level detector or $N-1$ two-level
atoms}, \pra~\textbf{86}, 015801 (2012).

\bibitem{Dodonov12b}
A. V. Dodonov and V. V. Dodonov,   \Title{Dynamical {C}asimir
effect in a cavity in the presence of a three-level atom},
\pra~\textbf{85}, 063804 (2012).

\bibitem{Dodonov97}
V. V. Dodonov and M. A. Andreata, \Title{Squeezing and photon
distribution in a vibrating cavity}, J. Phys. A: Math. Gen.
\textbf{32}, 6711 (1999).

\bibitem{Dodonov00}
V. V. Dodonov, A. B. Klimov, and V. I. Man'ko, \Title{Generation
of squeezed states in a resonator with a moving wall,} Phys. Lett.
A \textbf{149}, 225 (1990).

\bibitem{liuphonon}
Y. X. Liu, A. Miranowicz, Y. B. Gao, J. Bajer, C. P. Sun, and F.
Nori, \Title{Qubit-induced phonon blockade as a signature of
quantum behavior in nanomechanical resonators} \pra~\textbf{82},
032101 (2010).

\bibitem{Hartmann}
A. Ridolfo, M. Leib, S. Savasta, and M. J. Hartmann, \Title{Photon
blockade in the ultrastrong coupling regime,} \prl~\textbf{109},
193602 (2012).

\bibitem{Kossakowski72}
A. Kossakowski,   \Title{On quantum statistical mechanics of non-
Hamiltonian systems}, Rep. Math. Phys. {\bf 3}, 247 (1972).

\bibitem{Kossakowski76}
V. Gorini, A. Kossakowski, and E. C. G. Sudarshan,
\Title{Completely positive dynamical semigroups of $N$-level
systems}, J. Math. Phys. {\bf 17}, 821 (1976).

\bibitem{Kossakowski78}
V. Gorini, A. Frigerio, M. Verri, A. Kossakowski, and E. C. G.
Sudarshan, \Title{Properties of quantum {M}arkovian master
equations,} Rep. Math. Phys. {\bf 13}, 149 (1978).

\bibitem{Tan}
S. M. Tan, \Title{A computational toolbox for quantum and atomic
optics,} J. Opt. B~\textbf{1}, 424 (1999).

\bibitem{xunwei2}
X.-W. Xu and Y.-J. Li, \Title{Antibunching photons in a cavity
coupled to an optomechanical system,} J. Phys. B~\textbf{46},
035502 (2013).

\bibitem{Bamba}
M. Bamba, A. Imam$\breve{o}$glu, I. Carusotto, and C. Ciuti,
\Title{Origin of strong photon antibunching in weakly nonlinear
photonic molecules,} \pra~\textbf{83}, 021802(R) (2011).

\bibitem{Savona}
V. Savona, \Title{Unconventional photon blockade in coupled
optomechanical systems,} arXiv:1302.5937.

\bibitem{xunwei1} X. W. Xu, Y. J. Li, and Y. X. Liu,
\Title{Photon-induced tunneling in optomechanical systems,}
\pra~\textbf{87}, 025803 (2013).

\bibitem{Roque}
T. F. Roque and A. Vidiella-Barranco, \Title{Coherence properties
of coupled optomechanical cavities,} \josab~\textbf{31}, 1232
(2014).

\end{thebibliography}
\end{document}